\documentclass{ectj}
\received{December 2009}
\volume{10}
\title[Bridge volatility estimators]{Homogeneous Volatility Bridge Estimators}
\author[A. Saichev, D. Sornette, V. Filimonov, F. Corsi]{A. Saichev$^{1,3}$, D. Sornette$^{1,2}$, V. Filimonov$^{1,3}$ and F. Corsi$^4$}
\address{$^{1}$ETH Zurich -- Department of Management, Technology and Economics, Switzerland.}

\address{$^{2}$Swiss Finance Institute, 40, Boulevard du Pont-d' Arve, Case Postale~3, 1211 Geneva 4, Switzerland.}

\address{$^{3}$Nizhni Novgorod State University -- Department of Mathematics, Russia.}

\address{$^{4}$ University of Lugano and Swiss Finance Institute, Via G. Buffi 13,
CH-6904 Lugano, Switzerland. }
\email{dsornette@ethz.ch}

\usepackage{graphicx}
\usepackage{amsmath}
\usepackage{amssymb}
\usepackage{theorem}
\usepackage{url}

\newtheorem{theorem}{Theorem}

\newtheorem{remark}{Remark}

\newtheorem{definition}{Definition}

\newtheorem{lemma}{Lemma}

\begin{document}

\begin{abstract}
We present a theory of homogeneous volatility bridge estimators for
log-price stochastic processes. The main tool of our theory is the parsimonious encoding
of the information contained in the open, high and low prices of incomplete bridge, corresponding to given log-price stochastic process, and in its close value, for a given time interval.
The efficiency of the new proposed estimators is favorably compared with that of the Garman-Klass and Parkinson estimators.

\end{abstract}

\section{Introduction}

\setcounter{equation}{0}
\setcounter{theorem}{0}
\setcounter{remark}{0}
\setcounter{lemma}{0}
\setcounter{definition}{0}

Volatility, defined as the standard deviation of the increments of the log-price over a specific time interval, is a universally used risk indicator. With the growing availability of high-frequency tick-by-tick price time series,
a number of new efficient volatility estimators have been developed (see, for instance, Yang and Zhang (2000), Corsi et al. (2001), Andersen et al. (2003), A\"it-Sahalia (2005), Zhang et al. (2005)). However, for most
applications involving risk assessment and management of investment portfolios, it is the common
practice to use time series of prices recorded at fixed time intervals, such as 1 minute, 5 minutes,
30 minutes, 1 hour, 1 day, 1 week, 1 month and so on. For such time series, four prices are
actually recorded, called the open-high-low-close (OHLC) of the price for each time interval.

Our purpose is to provide new tools to exploit systematically the OHLC to improve
volatility estimators, compared with techniques using only the close price time series.
It is intuitively appealing that close-minus-open, high-minus-open and low-minus-open
should provide significant information on the variability of the price, that should help improve
the volatility estimators. We present here
a comprehensive theory of homogeneous volatility bridge estimators for
arbitrary stochastic processes, that fully exploit the OHLC prices.
For this, we have started to develop the theory of the most efficient point-wise homogeneous OHLC volatility estimators, valid for any price processes (Saichev et al., 2009). The main tool of our theory is the parsimonious encoding
of all the information contained in the mentioned OHLC prices for a given time interval in the form of general ``diagrams'' associated with the
joint distributions of the high-minus-open, low-minus-open and close-minus-open values. The diagrams
can be tailored to yield the most efficient estimators associated to any statistical properties of the underlying log-price stochastic process.

The present work extends and generalizes (Saichev et al., 2009) by developing most efficient
 OHLC bridge estimators. We find that the new OHLC bridge estimators are significantly more efficient than the OHLC estimators obtained from the untransformed (or ``unbridged'') process. For Wiener and similar
 processes, this can be intuitively understand as follows. It is well-known that the high and low values of a Wiener process are most probably found in
the neighborhood of the edges of the observation interval. In contrast, by construction of the bridge, its high and low values are in general distant from the edges. As a result, the high and low of a bridge incorporate significantly more information about the variability of the original stochastic process than its own high and low values.
This is the motivation for us to extend the theory of (Saichev et al., 2009) for bridges and to provide explicit analytical expressions for the most efficient point-wise volatility bridge estimators, based on the analytical expression of the
joint distribution of its high-minus-open, low-minus-open and close-minus-open values.

Our work also improves on the following papers as follows.  Garman and Klass (G\&K) (1980) introduced a quadratic estimator for the variance of the Wiener process for the log-price, which has rather low variance. Parkinson (PARK) (1980) proposed a simple quadratic variance estimator proportional to $(H-L)^2$, which is using only a part of the information available from  OHLC prices. Rogers and Satchell (R\&S) (1991,1994)
introduced another quadratic estimator for the variance of the Wiener process with drift, which is unbiased for all drifts.
Both G\&K and R\&S estimators are focused on the variance, and do not present estimators for the volatility, which is of obvious interest for financial applications. Yang and Zhang (2000)
produced an unbiased and efficient quadratic
variance estimator, taking into account the OHLC of log-prices for $n>1$ consecutive days. Their main novelty is to take into account the possible existence of jumps (or gaps) of prices from yesterday's close till today's open prices. Their minimization of the variance of their estimators requires the estimation of expectations of a quadratic
form of the OHLC which they only partly achieve due to the lack of knowledge of the full joint distribution, which we offer in this paper.
Chan and Lien (2003) compared the empirical effectiveness of four estimators, the PARK, the G\&K and R\&S ones,
and the naive excursion range $H-L$ estimator. From the perspective offered
by these previous works, the present paper can be viewed as providing their full underpinning theory,
since we are able to express efficient estimators in the presence of arbitrary constraints from the explicit knowledge of the joint distribution of the OHLC log-prices.

The paper is organized as follows. Section~2 describes the properties of the stochastic processes for which our theory of most efficient homogeneous volatility bridge estimators is developed. Section~3 derives the general expressions for the most efficient homogeneous volatility OHLC bridge estimators. Section~4 provides a detailed analytical description of the statistical properties of incomplete bridges of Wiener process with drift, describing log-price dynamics.  Section~5 compares the
efficiency of our derived most efficient homogeneous bridge estimators and the efficiency of the generalized G\&K bridge estimator and of the normalized PARK estimator. Section~6 tests our results using
synthetic time series generated by numerical simulations, which mimic the tick-by-tick nature of real
log-price processes. Section~7 concludes.

\section{Homogeneous volatility bridge estimators}

\setcounter{equation}{0}
\setcounter{theorem}{0}
\setcounter{remark}{0}
\setcounter{lemma}{0}
\setcounter{definition}{0}

The main goal of this paper is to construct efficient bridge estimators using the open, high, low, close (OHLC)
prices for the variance and the volatility of some asset log-price process $A(t)$.

\subsection{Volatility of order $\lambda$}

The conventional definition of the volatility $V(t_0,T_0)$ of a stochastic process $A(t)$
at time $t_0$ and time scale $T_0$ is the standard deviation of its increment
\[
\Delta(t_0,T_0) = A(t_0+T_0)-A(t_0)
\]
within the time interval $t\in(t_0,t_0+T_0)$:
\[
V(t_0,T_0) = \sqrt{\textrm{Var}\left[\Delta(t_0,T_0)\right]} .
\]
The time scale $T_0$ can be for instance 5 minutes, 1 day or 1 year, corresponding
respectively to intraday, daily or yearly volatility.

Since different measures of the variability of log-price processes are used
in the literature, it is convenient to define a generalized volatility of order $\lambda$ as follows.
\begin{definition}
\textnormal{The volatility of order $\lambda$ of the stochastic process $A(t)$ is the power $\lambda$ of
the conventional volatility
\[
V_\lambda(t_0,T_0) := V^\lambda(t_0,T_0) = \left(\textrm{Var}\left[\Delta(t_0,T_0)\right]\right)^{\lambda /2} .
\]}
\end{definition}

\begin{remark}
\textnormal{For $\lambda=1$, the volatility of order $\lambda$ coincides with the conventional volatility, while, for $\lambda=2$, $V_2(t_0,T_0)$ is the variance of the increment $\Delta(t_0,T_0)$. Most known estimators, for instance the R\&S, G\&K and PARK ones, are variance estimators. Introducing the volatility of order $\lambda$
gives us the possibility later on to compare the differences and relations between the volatility and variance estimators.}
\end{remark}

\subsection{Wiener process model of log-price increments}

We will analyze the properties of the estimators of the
volatility of order $\lambda$ for the Wiener process with drift,
posing without loss of generality $t_0=0$ and $A(0)=0$. This implies that
\begin{equation}\label{rtwpdrdef}
A(t) := \mu t + \sigma W(t) ,
\end{equation}
where $\mu$ is the drift of the log-price process $A(t)$ and $\sigma$ is its standard deviation at $t=1$, while $W(t)$ is the standard Wiener process with zero drift and
variance $\textrm{E}[W^2(t)]=t$.  The self-similar properties of the Wiener process allow us
to choose the time scale by $T_0=1$ without loss of generality, so that the
volatility of order $\lambda$ is simply equal to the standard deviation $\sigma$
raised to the power $\lambda$:
\[
V_\lambda := V_\lambda(t_0=0,T_0=1) = \sigma^\lambda .
\]
We analyze below the volatility estimators based on the high, low and close values of
the \emph{incomplete bridge} of the Wiener process with drift $A(t)$ defined by \eqref{rtwpdrdef}.
\begin{definition}
\textnormal{The stochastic process
\begin{equation}\label{bridgedeforig}
B(t,\kappa,T) := A(t) - \kappa {t \over T}  A(T) = \mu(1-\kappa) t + \sigma \left[W(t)-\kappa {t \over T} W(T) \right] ,
\end{equation}
where $\kappa$ is arbitrary constant, is called the incomplete bridge of
the original stochastic process $A(t)$. For $\kappa=1$, the incomplete bridge
is nothing but the standard (complete) bridge
\[
B(t,T) :=\sigma \left[W(t)- {t \over T} W(T) \right] .
\]
}
\end{definition}

Using the self-similar properties of the Wiener process,
one can rewrite \eqref{rtwpdrdef}, \eqref{bridgedeforig} in the form
\begin{equation}\label{bridgefactsigm}
A(t) = \sigma \sqrt{T} X\left({t \over T},\gamma\right) , \qquad
B(t,\kappa,T) = \sigma \sqrt{T} Y\left({t \over T},\kappa,\gamma\right),
\end{equation}
where $Y(t,\kappa,\gamma)$ is the incomplete bridge
\begin{equation}\label{inbrdef}
Y(t,\kappa,\gamma) := X(t,\gamma) - \kappa t X(1,\gamma)
\end{equation}
of the Wiener process with drift
\begin{equation}\label{wpdrdef}
X(t,\gamma) := \gamma t + W(t) , \qquad t\in(0,1) ,
\end{equation}
and the auxiliary parameter
\begin{equation}\label{gammadef}
\gamma = {\mu \over \sigma} \sqrt{T}
\end{equation}
plays the role of a 1$^{\rm st}$ standardized moment (or inverse coefficient of variation)
of the distribution of increments of the process $A(t)$ over the time interval $T$.
Figure~1 shows a realization of the Wiener process $W(t)$ and
its complete bridge. The high and low values of the
Wiener process and of its bridge are in general drastically different.

\begin{remark}
\textnormal{In financial markets applications, both the drift $\mu$ and the standard
deviation $\sigma$ are unknown. Thus, the value of the parameter $\gamma$ is unknown as well. Nevertheless, for the convenience of our analysis,  we will suppose in the following derivations that the value of parameter $\gamma$ is given. One can take into account the indeterminateness of the parameter $\gamma$ by exploring in detail the dependence as a function of $\gamma$ of the bias and of the efficiency of the OHLC volatility bridge estimators,
following the analysis performed by Saichev et al. (2009) for $\kappa =0$.
}
\end{remark}

\begin{figure}[h!]
\begin{center}
\includegraphics[width=0.7\textwidth]{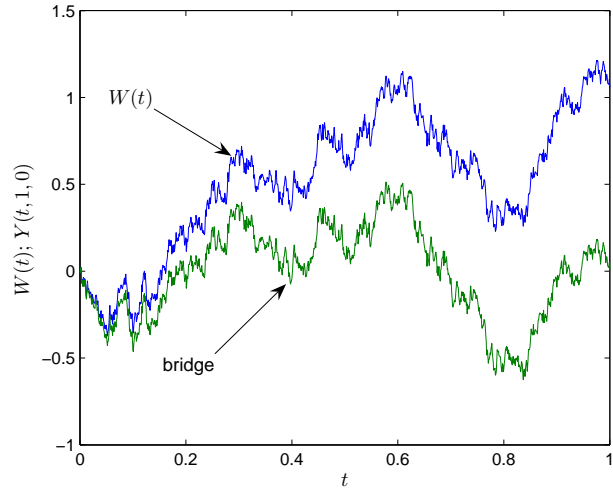}
\caption{A realization of the Wiener process $W(t)$ and complete bridge $W(t)-t W(1)$}
\end{center}
\end{figure}

\subsection{Homogeneous volatility bridge estimators}

\begin{definition}
\textnormal{A volatility estimator $V_\lambda$ is called an homogeneous OHLC volatility bridge estimator of order $\lambda$ if it has the form
\begin{equation}\label{volvarest}
\hat{\sigma}_\lambda = {1 \over T^{\lambda/2}} h_\lambda(\bar{H}, \bar{L}, \bar{C}) ,
\end{equation}
where $h_\lambda$ is a homogeneous function of order $\lambda$, the random values $\bar{H}$ and $\bar{L}$, are
the high and low of the incomplete bridge $B(t,\kappa,T)$ defined by \eqref{bridgedeforig} within the observation interval $(0,T)$,
\[
\bar{H} := \sup_{t\in(0,T)} B(t,\kappa,T) , \qquad \bar{L} := \inf_{t\in(0,T)} B(t,\kappa,T) ,
\]
and
\[
\bar{C}:=A(T) = \mu T + \sigma W(T)
\]
is the close value of the original stochastic process for the log-price $A(t)$.}
\end{definition}

A remarkable property of homogeneous estimators defined by \eqref{volvarest} is that, for a given $\gamma$, their statistical properties do not depend on the duration $T$ of the observation interval. Mathematically, this fact is expressed by the following theorem.
\begin{theorem}
The estimator defined by \eqref{volvarest} is equal to
\begin{equation}\label{eststhrucanonic}
\hat{\sigma}_\lambda = \sigma^\lambda h_\lambda(H,L,C) ,
\end{equation}
where $H$ and $L$ are the high and low values of the incomplete bridge $Y(t,\kappa,\gamma)$
defined by expression \eqref{inbrdef}
\begin{equation}\label{extremesydef}
H := \sup_{t\in(0,1)} Y(t,\kappa,\gamma) , \quad L := \inf_{t\in(0,1)} Y(t,\kappa,\gamma) ,
\end{equation}
while $C:=X(1,\gamma)$ is the close value of the Wiener process $X(t,\gamma)$ with drift,
defined by \eqref{wpdrdef}.
\end{theorem}

\emph{\textbf{Proof.}}
Substituting the right-hand-side of the equalities of \eqref{bridgefactsigm} into the right-hand-side of
expression \eqref{volvarest} and using the homogeneity of the function $h_\lambda$, we obtain equality \eqref{eststhrucanonic}. \hfill $\blacksquare$

\begin{definition}
\textnormal{We refer to the function
\begin{equation}\label{canestdef}
\hat{e}_\lambda = h_\lambda(H,L,C)
\end{equation}
as the canonical OHLC volatility bridge estimator of order $\lambda$. Using this definition, one can rewrite expression \eqref{eststhrucanonic} in the form
\begin{equation}\label{estscanonicform}
\hat{\sigma}_\lambda = \sigma^\lambda  \hat{e}_\lambda .
\end{equation}
}
\end{definition}

\begin{remark}
\textnormal{The statistical properties of the canonical estimators \eqref{canestdef} depend on the standard deviation $\sigma$ only through the parameter $\gamma$ defined in \eqref{gammadef}, which we assume in the following derivations to be known.}
\end{remark}

\section{Most efficient homogeneous bridge estimators}

\setcounter{equation}{0}
\setcounter{theorem}{0}
\setcounter{remark}{0}
\setcounter{lemma}{0}
\setcounter{definition}{0}

\subsection{Diagrams of homogeneous bridge estimators}

It results immediately from expressions \eqref{estscanonicform}, \eqref{canestdef}
that the homogeneous estimator given by \eqref{volvarest} is unbiased, if the expected value of
the corresponding canonical estimator given by \eqref{canestdef} is equal to unity:
\begin{equation}\label{unbiasdef}
\textrm{E}[\hat{e}_\lambda]=\textrm{E}[h_\lambda(H,L,C)] = 1 .
\end{equation}

\begin{definition}
\textnormal{The homogeneous volatility bridge estimator of order $\lambda$ given by \eqref{volvarest} is called
the most efficient one, for a given value $\gamma_0$ of the parameter $\gamma$ and for a fixed value of the parameter $\kappa$, if, for $\gamma=\gamma_0$ and fixed values of $\kappa$ and $\lambda$, the equality \eqref{unbiasdef} holds while the variance of the corresponding canonical estimator achieves the minimal value among the variances of all canonical estimators of given order $\lambda$ and for the same parameters $\gamma=\gamma_0$ and $\kappa$.}
\end{definition}

In this section, we provide the explicit expressions of the most efficient homogeneous volatility estimators.
For this, it is convenient to use a change of variables from the random variables $\{H,L,C\}$
to their corresponding spherical (geographic) coordinate variables $\{R, \Theta, \Phi\}$:
\begin{equation}\label{hlcthrurtph}
H = R \cos\Theta \cos\Phi , \quad L = R \cos\Theta \sin\Phi , \quad C = R \sin\Theta .
\end{equation}
Inversely, we have
\begin{equation}\label{geogrback}
R= \sqrt{\mathstrut H^2 +L^2 + C^2} , \quad
\Theta = \arctan\left({C \over \sqrt{\mathstrut H^2 + L^2}}\right) , \quad \Phi = \arctan\left({L \over H} \right).
\end{equation}
Substituting \eqref{hlcthrurtph} into \eqref{canestdef} and taking into account the homogeneity of
the function $h_\lambda$, we obtain
\begin{equation}\label{hatsthrediagr}
\hat{e}_\lambda = R^\lambda~ \psi_\lambda(\Theta, \Phi) ,
\end{equation}
where
\begin{equation}\label{diargdef}
\psi_\lambda(\theta, \phi) = h_\lambda(\cos\theta \cos\phi,\cos\theta \sin\phi,\sin\theta) .
\end{equation}

\begin{definition}
\textnormal{The function $\psi_\lambda(\theta,\phi)$ defined by expression \eqref{diargdef} is called the diagram of the canonical estimator of order $\lambda$.}
\end{definition}

\begin{remark}
\textnormal{The spherical coordinate system is intrinsic to
homogeneous estimators, allowing us to split them into a
known power function $R$ and
an arbitrary function of the variables $\Theta$ and $\Phi$ (see Eq. (\ref{hatsthrediagr})).
The spherical coordinate system reduces the search of efficient OHLC estimators
from three-dimensional functions to the appropriate
two-dimensional function $\psi_\lambda(\theta, \phi)$.}
\end{remark}

\subsection{Domain of possible $\{\Theta,\Phi\}$ values}

Below, we will need the domain of existence for the values of the random variables $\{R,\Theta,\Phi\}$
defined by \eqref{geogrback}. First, it is obvious that $R\in (0,\infty)$.
The domain $\mathcal{S}$ of the possible values of the two other random variables $\{\Theta,\Phi\}$
depends on the interplay between the random high $H$ and low $L$ of the incomplete bridge
given by \eqref{inbrdef}, and the close value $C$ of the Wiener process with drift
defined by \eqref{wpdrdef}. It will be clear below that $\mathcal{S}$ depends on the parameter $\kappa$. Thus, we denote it by $\mathcal{S}_\kappa$. We use the same notation $\mathcal{S}_\kappa$ for the domain of the arguments $\{\theta,\phi\}$ of the diagram $\psi_\lambda(\theta,\phi)$ defined by \eqref{diargdef}: $\{\theta,\phi\}\in \mathcal{S}_\kappa$. Since $H\geqslant 0$ and $L\leqslant 0$, in view of \eqref{hlcthrurtph}, we have
\[
\tan\Phi = {L \over H} \in (-\infty,0] \qquad \Rightarrow \qquad  -{\pi \over 2} \leqslant \Phi < 0  .
\]
In turn, as it seen from \eqref{inbrdef} and \eqref{wpdrdef}, the values $\{H,L,C\}$ satisfy to the inequalities $L \leqslant (1-\kappa)C \leqslant H$ or, using \eqref{hlcthrurtph},
\[
\sin\Phi \leqslant (1-\kappa)\tan\Theta  \leqslant \cos\Phi \, \Rightarrow \, \arctan\left({\sin\Phi\over 1-\kappa}\right) \leqslant \Theta \leqslant\arctan\left({\cos\Phi\over 1-\kappa}\right)  .
\]
Thus
\begin{equation}
\mathcal{S}_\kappa = \left\{\arctan\left({\sin\Phi\over 1-\kappa}\right) \leqslant \Theta \leqslant\arctan\left({\cos\Phi\over 1-\kappa}\right), -{\pi \over 2} \leqslant \Phi < 0 \right\} .
\label{thyjumldls}
\end{equation}

\subsection{Most efficient OHLC homogeneous bridge estimators}

Let us denote the joint probability density function (pdf) of
the random variables $\{H,L,C\}$ by $Q(h,\ell,c;\kappa,\gamma)$.
Then, the expected value of the canonical estimator defined by \eqref{hatsthrediagr} is equal to
\begin{equation}\label{expvalcanes}
\textrm{E}[\hat{e}_\lambda|\kappa,\gamma] = \mathcal{M}_\lambda(\kappa,\gamma):= \iint\limits_{\mathcal{S}_\kappa} \psi_\lambda(\theta,\phi) ~ g_\lambda(\theta,\phi;\kappa,\gamma) \cos \theta d\theta d\phi,
\end{equation}
where
\begin{equation}\label{glambdef}
g_\lambda(\theta,\phi; \kappa, \gamma) = \int_0^\infty \rho^{\lambda+2} Q(\rho\cos \theta \cos\phi, \rho \cos\theta\sin\phi,\rho \sin\theta;\kappa,\gamma) d\rho .
\end{equation}
Accordingly, at $\gamma=\gamma_0$ and given $\kappa$, one can represent the diagram of any unbiased, homogeneous estimator in the form
\begin{equation}\label{unbiasdiagr}
\psi_\lambda(\theta,\phi;\kappa,\gamma_0) = {G(\theta,\phi) \over \iint\limits_{\mathcal{S}_\kappa} G(\theta,\phi) ~ g_\lambda(\theta,\phi;\kappa,\gamma_0) \cos \theta d\theta d\phi} ,
\end{equation}
where $G(\theta,\phi)$ is an arbitrary function.
The following theorem gives the expression for the diagram \eqref{unbiasdiagr} corresponding
to the most efficient (for any given $\kappa$ and $\gamma=\gamma_0$) homogeneous
estimator of order $\lambda$.
\begin{theorem}\label{mostefth}
The diagram of the most efficient (for a given $\kappa$ and $\gamma=\gamma_0$) homogeneous bridge estimator of order $\lambda$ is equal to
\begin{equation}\label{diagrmostefdef}
\psi_{\textrm{me},\lambda}(\theta,\phi;\kappa, \gamma_0) = { G_\lambda(\theta,\phi;\kappa, \gamma_0) \over \mathcal{E}_\lambda(\kappa,\gamma_0)}  , \qquad \{\theta,\phi\} \in \mathcal{S}_\kappa  ,
\end{equation}
where
\begin{equation}\label{mathedef}
G_\lambda(\theta,\phi;\kappa, \gamma)= {g_\lambda(\theta,\phi;\kappa,\gamma) \over g_{2\lambda}(\theta,\phi;\kappa, \gamma)} , \quad
\mathcal{E}_\lambda(\kappa,\gamma) = \iint\limits_{\mathcal{S}_\kappa}
{g_\lambda^2(\theta,\phi;\kappa,\gamma) \over g_{2\lambda}(\theta,\phi;\kappa, \gamma)}\cos\theta d\theta d\phi .
\end{equation}
\end{theorem}

 The proof is given in Appendix \ref{thyh3q}.

\section{Statistical description of incomplete bridges}

\setcounter{equation}{0}
\setcounter{theorem}{0}
\setcounter{remark}{0}
\setcounter{lemma}{0}
\setcounter{definition}{0}

\subsection{Identical in law Wiener process}

In order to get the most efficient homogeneous OHLC bridge estimator, we need the pdf $Q(h,\ell,c;\kappa,\gamma)$ of the high and low of the incomplete bridge $Y(t,\kappa,\gamma)$ defined by \eqref{inbrdef} and the close value $C$ of the underlying process $X(t,\gamma)$ defined by \eqref{wpdrdef}. Before giving the
explicit solution, it is useful to discuss their general statistical properties.

\begin{theorem}\label{identlawtheorem}
The incomplete bridge $Y(t,\kappa,\gamma)$ given by \eqref{inbrdef} is identical in law to the diffusion process
\begin{equation}\label{yexplmathcal}
\mathcal{Y}(t,\kappa,\gamma) := \gamma (1-\kappa) t +\mathcal{W}(t,\kappa) ,
\end{equation}
where
\begin{equation}\label{bomeganew}
\mathcal{W}(t,\kappa) := (1-t+(1-\kappa)^2 t) W\left( {t \over 1-t+(1-\kappa)^2 t} \right) .
\end{equation}
\end{theorem}

\emph{\textbf{Proof.}} After substitution \eqref{wpdrdef} into \eqref{inbrdef}, we obtain
\begin{equation}\label{yexplomega}
Y(t,\kappa,\gamma) = \gamma (1-\kappa) t +\Omega(t,\kappa) ,
\end{equation}
where
\[
\Omega(t,\kappa) := W(t) -\kappa t W(1) .
\]
One can easily verify that $\Omega(t,\kappa)$ is a Gaussian process with zero mean and covariance
given by
\begin{equation}\label{bomegacov}
\textrm{E}[\Omega(t_1,\kappa)\Omega(t_2,\kappa)] = (t_1 \wedge t_2) - [1- (1-\kappa)^2] t_1 t_2 , \quad 0\leqslant t_1, t_2 \leqslant 1 .
\end{equation}
Direct calculations show that the Gaussian process
$\mathcal{W}(t,\kappa)$ defined by \eqref{bomeganew} is also characterized by a zero mean and the same covariance \eqref{bomegacov}. This implies that the incomplete bridge $Y(t,\kappa,\gamma)$ given by \eqref{yexplomega} is identical in law to the diffusion process $\mathcal{Y}(t,\kappa,\gamma)$ defined in \eqref{yexplmathcal}. \hfill $\blacksquare$

\subsection{Change of time \label{yheyhbwgr}}

Henceforth, for the analysis of the statistical properties of the incomplete bridge $Y(t,\kappa,\gamma)$
defined by \eqref{inbrdef}, we will use the equivalence in law stated in theorem~\ref{identlawtheorem},
which allows us to
to replace the incomplete bridge by the diffusion process $\mathcal{Y}(t,\kappa,\gamma)$ defined by \eqref{yexplmathcal}. As will be clear below, it is convenient to explore the extremal properties of the diffusion process $\mathcal{Y}(t,\kappa,\gamma)$ using the change of time
\[
\tau = \tau(t,\kappa):= {(1-\kappa)^2 t \over 1-t+(1-\kappa)^2 t} .
\]
Inversely,
\[
t =t(\tau,\kappa):= {\tau \over \tau + (1-\kappa)^2 (1-\tau)}  .
\]
The function $\tau(t,\kappa)$ maps the interval $t\in(0,1)$ onto the same interval $\tau\in(0,1)$.

Let us introduce the auxiliary stochastic process
\begin{equation}\label{ztaudef}
\mathcal{Z}(\tau,\kappa,\gamma) := \mathcal{Y}(t(\tau,\kappa),\kappa,\gamma) .
\end{equation}
Using relations \eqref{yexplmathcal}, \eqref{bomeganew}, and self-similar properties of Wiener process, rewrite $\mathcal{Z}(\tau,\kappa,\gamma)$ in the form
\begin{equation}\label{ztauexplform}
\mathcal{Z}(\tau,\kappa,\gamma) = {1-\kappa  \over \tau+ (1-\kappa)^2(1- \tau)} \left[\gamma \tau + W(\tau)\right]  .
\end{equation}
Below we assume, for definiteness, $\kappa<1$.

It follows from the construction \eqref{ztaudef} of the stochastic process $\mathcal{Z}(\tau,\kappa,\gamma)$ and from the equality \eqref{ztauexplform} that the following inequalities are equivalent
\begin{equation}\label{yzextrequiv}
L \leqslant \mathcal{Y}(t,\kappa,\gamma) \leqslant H  , \quad \Leftrightarrow \quad
a + \alpha \tau \leqslant W(\tau) \leqslant b + \beta \tau , \quad t, \tau\in(0,1) ,
\end{equation}
where
\begin{equation}\label{albetdef}
a = (1-\kappa) L , ~ b = (1-\kappa) H , ~
\alpha = {1 -(1-\kappa)^2 \over 1-\kappa} L -\gamma ,  ~ \beta = {1 -(1-\kappa)^2 \over 1-\kappa} H -\gamma .
\end{equation}

Additionally, the close value $C=X(1,\gamma)$ of the stochastic process $X(t,\gamma)$ \eqref{wpdrdef}
is tied to the close value of the incomplete bridge $Y(t,\kappa,\gamma)$ given by \eqref{inbrdef} by the equality
$Y(1,\kappa,\gamma) = (1-\kappa)C$.
In turn, it follows from the identity in law of the stochastic processes $Y(t,\kappa,\gamma)$ and $\mathcal{Y}(t,\kappa,\gamma)$ and from relations \eqref{ztaudef}, \eqref{ztauexplform} that one may replace $Y(1,\kappa,\gamma)$ by
\[
\mathcal{Z}(1,\kappa,\gamma) = (1-\kappa) [\gamma + W(1)] .
\]
Thus, one obtains
\begin{equation}\label{czoneomeq}
W(1) = C-\gamma  .
\end{equation}

\subsection{Diffusion equation}

Let us define the probability
\[
f(h,\ell,c;\kappa,\gamma)dc :=\Pr\{C\in(c,c+dc)\cap \ell\leqslant Y(t,\kappa,\gamma)\leqslant h; t\in(0,1)\} .
\]
Then, the  joint pdf of the high and low values $\{H,L\}$ \eqref{extremesydef} of the incomplete bridge $Y(t,\kappa,\gamma)$, and of the close value $C$ of the original process $X(t,\gamma)$ \eqref{wpdrdef}, is equal to
\begin{equation}\label{qfpart}
Q(h,\ell,c;\kappa,\gamma) = -{\partial^2 f(h,\ell,c;\kappa,\gamma) \over \partial h\partial \ell} ,
\end{equation}
\[
h>h_-, ~ \ell<\ell_+, ~ {\ell \over 1-\kappa} \leqslant c \leqslant {h \over 1-\kappa},~
h_- = 0 \vee (1-\kappa) c, ~ \ell_+ = 0 \wedge (1-\kappa) c.
\]
From the relations of the previous subsection \ref{yheyhbwgr}, one can express the function $f(h,\ell,c;\kappa,\gamma)$ via the auxiliary function $\varphi(\omega;\tau)$
\[
\varphi(\omega;\tau) d\omega :=
\Pr\{W(\tau)\in(\omega,\omega+d\omega) \cap a +\alpha \tau' \leqslant W(\tau')\leqslant b+ \beta\tau'; \tau'\in(0,\tau)\} ,
\]
according to
\begin{equation}\label{fthruphi}
f(h,\ell,c;\kappa,\gamma) = \varphi(c-\gamma;1,a, b,\alpha,\beta) .
\end{equation}

The theory of Wiener processes implies that the auxiliary function  $\varphi(\omega;\tau)$
is the solution of the diffusion equation
\begin{equation}\label{difeqphi}
{\partial \varphi \over \partial \tau} = {1 \over 2} {\partial^2 \varphi \over \partial \omega^2} ,
\end{equation}
with initial condition
\begin{equation}\label{phiincond}
\varphi(\omega;\tau=0) = \delta(\omega)
\end{equation}
and absorbing boundary conditions
\begin{equation}\label{absboundcond}
\varphi(\omega=a+\alpha \tau;\tau)= 0 , \quad \varphi(\omega=b+\beta\tau;\tau)= 0 , \quad \tau>0 ,
\end{equation}
which account for the inequalities \eqref{yzextrequiv}.

Below, we solve this initial-boundary problem  \eqref{difeqphi}, \eqref{phiincond}, \eqref{absboundcond} and
determine the joint pdf of the high and low values of the incomplete bridge
$Y(t,\kappa,\gamma)$ and of the close value of the Wiener process $X(t,\gamma)$ with drift, using the
following relation that derives from  \eqref{qfpart} and \eqref{fthruphi}:
\begin{equation}\label{jpdfthruphi}
Q(h,\ell,c;\kappa,\gamma) = -{\partial^2 \varphi(c-\gamma;1,a, b,\alpha,\beta) \over \partial h\partial \ell} .
\end{equation}

\subsection{Useful properties of the solutions of diffusion equations}

Before solving explicitly the initial-boundary problem \eqref{difeqphi}, \eqref{phiincond}, \eqref{absboundcond}, it is useful to present some general properties of its solutions. Firstly, if $\varphi(\omega;\tau)$ is some solution of diffusion equation \eqref{difeqphi}, then $A\varphi(\omega+a;\tau)$, where $a$ and $A$ are arbitrary constants, is also a solution. Such relation tying together different solutions of the same diffusion equation can be written as
\begin{equation}\label{shiftrel}
\varphi(\omega;\tau) \qquad \longleftrightarrow \qquad A \varphi(\omega+a;\tau)  .
\end{equation}
In order to solve the initial-boundary problem  \eqref{difeqphi}, \eqref{phiincond}, \eqref{absboundcond},
we will need two lemmas.
\begin{lemma}\label{lemone}
If $\varphi(\omega,\tau)$ of the form
\begin{equation}\label{soldifinconphi}
\varphi(\omega;\tau) = {1 \over \sqrt{2\pi \tau}} \int_{-\infty}^\infty \varphi(y)
\exp\left( - {(\omega-y)^2 \over 2\tau} \right) dy
\end{equation}
is a solution of the diffusion equation \eqref{difeqphi}, satisfying the initial condition
\[
\varphi(\omega;t) = \varphi(\omega) ,
\]
where $\varphi(\omega)$ is such that $\varphi(\omega,\tau)$ is a continuous function of $\omega$ for any $\tau>0$, then it generates a family of continuous solutions via the transformation
\begin{equation}\label{rrelrile}
\varphi(\omega;\tau) \qquad \longleftrightarrow \qquad A \varphi(2 \alpha \tau -\omega;\tau)~ e^{2\alpha(\alpha \tau-\omega)} ,
\end{equation}
where $A$ and $\alpha$ are arbitrary constants.
\end{lemma}

\emph{\textbf{Proof.}}
Let us write the function on the right of the relation \eqref{rrelrile} in explicit form:
\[
\varphi(2 \alpha \tau -\omega;\tau)~ e^{2\alpha(\alpha \tau-\omega)} =
{1 \over \sqrt{2\pi \tau}} \int_{-\infty}^\infty \varphi(y) \exp\left( - {(2\alpha \tau - \omega-y)^2 \over 2\tau} \right) dy ~ e^{2\alpha(\alpha \tau-\omega)} .
\]
Since
\[
-{(2\alpha \tau - \omega-y)^2 \over 2\tau} + 2\alpha(\alpha \tau-\omega) =  - {(\omega+y)^2 \over 2\tau} + 2\alpha y ,
\]
the right-hand side of relation \eqref{rrelrile} is a continuous solution of the diffusion equation \eqref{difeqphi}, satisfying the initial condition
\[
\tilde{\varphi}(\omega) = \varphi(-\omega) ~ e^{2\alpha \omega} ,
\]
analogously to \eqref{soldifinconphi}. \hfill $\blacksquare$

The second lemma needed to find the solution of the initial-boundary problem  \eqref{difeqphi}, \eqref{phiincond}, \eqref{absboundcond} can be stated as follows.

\begin{lemma}\label{lem2}
Consider the function $\varphi(\omega)$ which verifies  to symmetry relation
\begin{equation}\label{symrel}
\varphi(\omega) = - \varphi(2 a - \omega)~ e^{2\alpha (a-\omega)} .
\end{equation}
Then, the solution $\varphi(\omega;\tau)$ of the diffusion equation \eqref{soldifinconphi},
which is continuous with respect to $\omega$ and with initial condition equal to
$\varphi(\omega)$, is vanishing on the line $\omega = a + \alpha \tau$:
\[
\varphi(a+\alpha \tau;\tau) = 0 , \qquad \tau>0 .
\]
\end{lemma}

\emph{\textbf{Proof.}}
Consider the function
\begin{equation}\label{tildereverse}
\tilde{\varphi}(\omega;\tau) = \varphi(2 \alpha \tau+ 2a -\omega;\tau)~ e^{2\alpha(\alpha \tau+a-\omega)} ,
\end{equation}
where $\varphi(\omega;\tau)$ is given by expression \eqref{soldifinconphi} and $\varphi(\omega)$ obeys to symmetry relation \eqref{symrel}. It follows from \eqref{shiftrel}, \eqref{rrelrile} and from the conditions of lemma~\ref{lemone}, that $\tilde{\varphi}(\omega;\tau)$ satisfies the diffusion equation \eqref{difeqphi} and is, for $\tau>0$, a continuous function of the argument $\omega$.
Expressions \eqref{tildereverse} and \eqref{symrel} ensure that the solution $\tilde{\varphi}(\omega;\tau)$ satisfies the initial condition
\[
\tilde{\varphi}(\omega;\tau=0) = \varphi(2 a - \omega)~ e^{2\alpha (a-\omega)} = -\varphi(\omega) .
\]
This means in turn that
\[
\tilde{\varphi}(\omega;\tau) = -\varphi(\omega;\tau) ,
\]
or in explicit form
\[
\varphi(\omega;\tau) = -\varphi(2 \alpha \tau+ 2a -\omega;\tau)~ e^{2\alpha(\alpha \tau+a-\omega)} .
\]
In particular
\[
\varphi(a+\alpha \tau;\tau) = - \varphi(a+\alpha \tau;\tau) \quad \Rightarrow \quad \varphi(a+\alpha \tau;\tau)= 0, \quad \tau>0 . \eqno{\blacksquare}
\]

\subsection{Solution of the initial-boundary problem}

The solution of the initial-boundary problem  \eqref{difeqphi}, \eqref{phiincond}, \eqref{absboundcond}
is obtained below by using the reflection method and the final result is stated in the following theorem.
\begin{theorem} \label{jujuouyyn}
The solution of the diffusion equation \eqref{difeqphi}, satisfying the initial-boundary conditions \eqref{phiincond}, \eqref{absboundcond}, is given by
\begin{eqnarray}\label{solbidifprob}
\varphi(\omega;\tau) = \sum_{m=-\infty}^\infty e^{2(\alpha-\beta) (b-a) m^2+2 (\alpha b-\beta a) m} \times
\\
\left[g(\omega+2 m(b-a);\tau)- e^{2a(2(\beta -\alpha) m-\alpha)} g(\omega+2m(b-a)-2a;\tau) \right] ,  \nonumber
\end{eqnarray}
where
\[
g(\omega;\tau) = {1 \over \sqrt{2 \pi \tau}} \exp\left(-{\omega^2 \over 2 \tau} \right).
\]
\end{theorem}

The proof is given in Appendix \ref{thyh3qetgq}.

Substituting \eqref{solbidifprob} with \eqref{albetdef} into \eqref{fthruphi}, we obtain
\begin{equation}\label{fhlcexplis}
f(h,\ell,c; \kappa,\gamma) = g(c-\gamma)
\sum_{m=-\infty}^\infty e^{-2 (h-\ell)^2 m^2 -2 m (h-\ell) (1-\kappa) c}
\left[1 - e^{4 (h-\ell) \ell m-2 \ell(\ell- (1-\kappa) c)} \right] ,
\end{equation}
where
\[
g(c) ={1 \over \sqrt{2\pi}} \exp\left( -{c^2 \over 2}\right) .
\]

\subsection{Joint pdf of high, low and close values}

Using relations \eqref{fthruphi}, \eqref{jpdfthruphi} and \eqref{fhlcexplis}, we obtain the sought joint pdf $Q(h,\ell,c;\kappa,\gamma)$ of the high and low values $\{H,L\}$ defined by \eqref{extremesydef} of the incomplete bridge $Y(t,\kappa,\gamma)$ defined by \eqref{inbrdef}, together with the close value $C=X(1,\gamma)$ of the Wiener process $X(t,\gamma)$ with drift given by \eqref{wpdrdef}. Namely,
\begin{equation}\label{qhlcpdf}
Q(h,\ell,c;\kappa,\gamma) =  g(c-\gamma) \mathcal{R}(h,\ell;\kappa| c) ,
\end{equation}
where
\begin{equation}\label{mathrdef}
\mathcal{R}(h,\ell;\kappa| c) = \sum_{m=-\infty}^\infty m
\big[ m \mathcal{D}(m(h-\ell), (1-\kappa)c) + (1-m) \mathcal{D}(m(h-\ell)+\ell,(1-\kappa) c)\big]
\end{equation}
and
\begin{equation}\label{mathddef}
\mathcal{D}(h,c) = 4 [(c-2h)^2-1] e^{2 h(c-h)} .
\end{equation}
Obviously, $\mathcal{R}(h,\ell;\kappa| c)$ is the conditional pdf of the high and low values $\{H,L\}$, under
the condition that the close value $C$ is equal to a given $c$. For any $\kappa$,
the conditional pdf $\mathcal{R}(h,\ell;\kappa| c)$ does not depend on
the normalized drift parameter $\gamma$. Furthermore, it satisfies the normalizing condition
\[
\int_{h_-}^\infty dh \int^{\ell_+}_{-\infty} d\ell ~ \mathcal{R}(h,\ell;\kappa| c) = 1 .
\]

Taking the limit $\kappa\to1$ corresponds to the complete bridge, which
is an important case for our analysis below. Let us thus define
the joint pdf limit
\[
Q(h,\ell,c;\gamma) := \lim_{\kappa\to 1} Q(h,\ell,c;\kappa,\gamma) .
\]
Expressions \eqref{wpdrdef}, \eqref{mathrdef} and \eqref{mathddef} show that it is equal to
\begin{equation}\label{qpdfktoone}
Q(h,\ell,c;\gamma) = g(c-\gamma) \mathcal{R}(h,\ell) , \qquad
-\infty < c < \infty , \quad h>0 , \qquad \ell < 0 ,
\end{equation}
where
\begin{equation}\label{mathrhldef}
\mathcal{R}(h,\ell) =
\sum_{m=-\infty}^\infty m
\left[ m \mathcal{D}(m(h-\ell)) + (1-m) \mathcal{D}(m(h-\ell)+\ell)\right]
\end{equation}
and
\[
\mathcal{D}(h) = 4 (4 h^2 - 1)~ e^{-2 h^2} .
\]

Expression \eqref{qpdfktoone} has a clear probabilistic interpretation. It means that the
high and low values $\{H,L\}$ of the complete bridge $Y(t,1,\gamma)$ are statistically independent from
the close value $C=X(1,\gamma)$ of the original Wiener process with drift. Accordingly,
$\mathcal{R}(h,\ell)$ given by \eqref{mathrhldef} reduces to the unconditional joint pdf of the high and low values of the complete bridge.

\subsection{Diagrams of the most efficient homogeneous OHLC bridge estimators}

Knowing the joint pdf of the random variables $\{H,L,C\}$, one can calculate
the auxiliary functions $g_\lambda(\theta,\phi;\kappa,\gamma)$ \eqref{glambdef}
needed in the definition of the diagrams \eqref{diagrmostefdef}, \eqref{mathedef} of the most efficient homogeneous OHLC volatility bridge estimators. This allows us to derive a number of properties of the
functions $g_\lambda(\theta,\phi;\kappa,\gamma)$. It follows from \eqref{glambdef} and \eqref{qhlcpdf}, \eqref{mathrdef}, that
\begin{eqnarray}\label{glasum}
g_\lambda(\theta,\phi;\kappa,\gamma) & = &
\\
{1 \over \sqrt{2\pi}} e^{-\gamma^2/ 2} \sum_{m=-\infty}^\infty m
\big[ m I_\lambda(m (\tilde{h}-\tilde{l}),\tilde{c};\kappa,\gamma) & + & (1-m) I_\lambda(m(\tilde{h}-\tilde{l})+\tilde{l},\tilde{c};\kappa,\gamma) \big] , \nonumber
\end{eqnarray}
where
\[
I_\lambda(h,c;\kappa,\gamma) = \int_0^\infty \rho^{2+\lambda} \exp\left(\gamma c\rho-{c^2 \over 2}\rho^2\right) \mathcal{D}(h\rho,(1-\kappa) c\rho) d\rho ,
\]
and
\[
\tilde{h}= \cos\theta \cos\phi , \qquad \tilde{l} = \cos\theta\sin\phi , \qquad \tilde{c} = \sin\theta .
\]

All calculations done, the explicit expression of $I_\lambda(h,c;\kappa, \gamma)$ reads
\begin{eqnarray}
I_\lambda(h,c;\kappa, \gamma) &=& \left({2 \over a} \right)^{3+{\lambda \over 2}} \times \nonumber
\\
\bigg[b \sqrt{2a}~ \Gamma\left({5+\lambda \over 2}\right) M\left({5+\lambda \over 2},{1 \over 2}, {d^2 \over 2a} \right)
&-& a\sqrt{{a \over 2}}~ \Gamma\left({3+\lambda \over 2}\right)M\left({3+\lambda \over 2}, {1 \over 2}, {d^2 \over 2 a}\right) \nonumber
\\
+ 2 d b~ \Gamma\left(3+{\lambda \over 2}\right) M\left(3+{\lambda \over 2},{3 \over 2}, {d^2 \over 2a} \right)
&-& d a~ \Gamma\left(2 +{\lambda \over 2}\right) M\left(2 +{\lambda \over 2},{3 \over 2}, {d^2 \over 2 a} \right) \bigg] . \nonumber
\end{eqnarray}
Here,
$$
M(a, b, z) :=  {\Gamma(b) \over \Gamma(a) \Gamma(b-a)} \int_0^1 du ~e^{zu} u^{a-1} (1-u)^{b-a-1}~,~~~~~~{\rm Re}\{b\} > {\rm Re}\{a\} >0
$$
is the Kummer function (see Abramowitz M., and A. Stegun. (1964)). We have used the following notations
\[
a = 4 h(h-(1-\kappa) c) + c^2  , \quad b = (2 h-(1-\kappa) c)^2 , \qquad d = \gamma c .
\]

In the particular case $\gamma=0$, we obtain
\begin{eqnarray}\label{ilafrac}
&& I_\lambda(h,c;\kappa):=I_\lambda(h,c;\kappa,\gamma=0) = \qquad \\
&& 2^{5+\lambda \over 2} \Gamma\left({3 + \lambda \over 2}\right) {(3+\lambda)[2h - (1-\kappa) c]^2-(2 h-c)^2  - 4 \kappa c h \over [(2 h-c)^2+4 h \kappa c]^{{5+\lambda \over 2}}} .
\nonumber
\end{eqnarray}

Figure~2 shows a 3D plot of the diagram obtained from \eqref{diagrmostefdef}, \eqref{mathedef}, \eqref{glasum}, \eqref{ilafrac} of the most efficient variance bridge estimator, for $\kappa=0.95$ and $\gamma=0$. On the plane $(\theta,\phi)$ depicted boundary of domain $\mathcal{S}_\kappa$ \eqref{thyjumldls}.

\begin{figure}[h!]
\begin{center}
\includegraphics[width=0.99\textwidth]{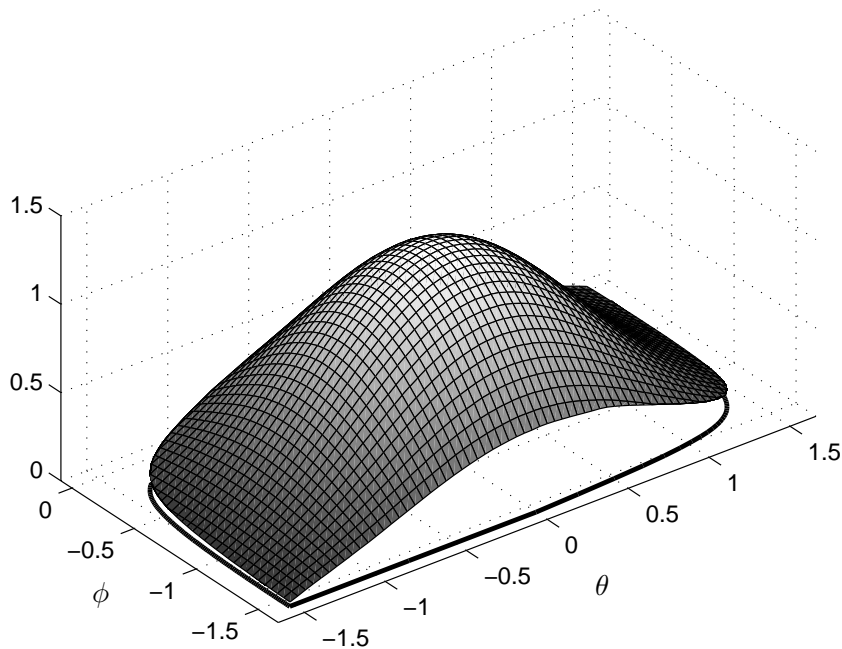}
\caption{Diagram of the most efficient variance estimator, for $\kappa=0.95$ and $\gamma=0$}
\end{center}
\end{figure}

\section{Comparison of the most efficient bridge estimators with the G\&K and PARK estimators}

\setcounter{equation}{0}
\setcounter{theorem}{0}
\setcounter{remark}{0}
\setcounter{lemma}{0}
\setcounter{definition}{0}

\subsection{Expectation and variance of arbitrary canonical bridge estimators}

In this section, we compare the efficiency of the most efficient homogeneous bridge estimators
derived in previous sections with that of the G\&K and PARK estimators.
We thus give the formulas for the expected value and the variance of
arbitrary canonical homogeneous OHLC bridge estimators defined by \eqref{hatsthrediagr}.
First, their expected values are $\mathcal{M}_\lambda(\kappa,\gamma)$ given by \eqref{expvalcanes}.

In general, homogeneous bridge estimators are biased. Thus, one needs a normalization procedure
for a practical comparison. We will normalize  the homogeneous bridge estimators
by the one obtained for a zero drift ($\gamma=0$). Thus, for each estimator \eqref{hatsthrediagr},
we consider its normalized version
\[
\tilde{e}_\lambda = R^\lambda { \psi_\lambda(\Theta, \Phi) \over \mathcal{M}_\lambda(\kappa)} ,
\]
where $\mathcal{M}_\lambda(\kappa) = \mathcal{M}(\kappa,\gamma= 0)$.
The expected value of the normalized estimator is
\[
\textrm{E}[\tilde{e}_\lambda|\kappa,\gamma] = {\mathcal{M}_\lambda(\kappa,\gamma) \over \mathcal{M}_\lambda(\kappa)} .
\]
The variance of the normalized (at $\gamma=0$) estimator is
\[
\textrm{Var}[\tilde{e}_\lambda|\kappa,\gamma] = { \mathcal{N}_\lambda(\kappa,\gamma) - \mathcal{M}^2_\lambda(\kappa,\gamma) \over \mathcal{M}^2_\lambda(\kappa)} ,
\]
where
\[
\mathcal{N}_\lambda(\kappa,\gamma) = \iint\limits_{\mathcal{S}_\kappa} \psi^2_\lambda(\theta,\phi) g_{2\lambda}(\theta,\phi; \kappa, \gamma) \cos\theta d\theta d\phi .
\]

\subsection{Generalized G\&K bridge estimator}

We recall that the G\&K canonical variance estimator is given by
\begin{eqnarray}\label{gkcanestdef}
\hat{e}_\textrm{GK} &=& k_1 (H-L)^2 - k_2 (C (H+L) -2 HL) -k_3 C^2 ,
\\
k_1 &=& 0.511 , \qquad k_2 = 0.019 , \qquad k_3 = 0.383 . \nonumber
\end{eqnarray}
The random variables $\{H,L,C\}$ are the high, low and close values of the Wiener process $X(t,\gamma)$ with drift defined by \eqref{wpdrdef}. In order to compare the efficiencies of the G\&K estimator and of most efficient bridge estimators, we modify the G\&K estimator \eqref{gkcanestdef} by replacing the high, low and close values of the Wiener process $X(t,\gamma)$ with drift by the high, low and close values of the incomplete bridge $Y(t,\kappa,\gamma)$ defined by \eqref{inbrdef}. This yields
\begin{equation}\label{gkbridgest}
\hat{e}_\textrm{GK}(\kappa) = k_1 (H-L)^2 - k_2 ((1-\kappa) C (H+L) -2 HL) -k_3 (1-\kappa)^2 C^2 .
\end{equation}
The estimator \eqref{gkbridgest} can be expressed in a form analogous to \eqref{hatsthrediagr},
\begin{equation}\label{gkdiagrdef}
\hat{e}_\textrm{GK} = R^2 \psi_\textrm{GK}(\Theta,\Phi, \kappa) ,
\end{equation}
with
\begin{eqnarray}
\psi_\textrm{GK}(\theta,\phi,\kappa) &=& k_1 \cos^2\theta (\cos\phi - \sin \phi)^2 \nonumber
\\
&+& k_2 \left[ \cos^2 \theta \sin 2\phi - {1-\kappa \over 2} \sin 2 \theta (\cos \phi  + \sin \phi) \right] - k_3 (1-\kappa)^2\sin^2 \theta . \nonumber
\end{eqnarray}

To  compare the efficiencies of the G\&K estimator and of the most efficient bridge estimators of arbitrary order $\lambda$, let us introduce the G\&K estimator of order $\lambda$:
\begin{equation}\label{gkmodifieddef}
\tilde{e}_{\textrm{GK},\lambda} = {R^\lambda \over
\mathcal{M}_{\textrm{GK},\lambda}(\kappa)}~ \psi_\textrm{GK}^{\lambda/2}(\Theta,\Phi,\kappa) ,
\end{equation}
where $\mathcal{M}_{\textrm{GK},\lambda}(\kappa)$ is given by the following expression
\[
\mathcal{M}_{\textrm{GK},\lambda}(\kappa) = \iint\limits_{\mathcal{S}_\kappa} \psi_\textrm{GK}^{\lambda/2}(\theta,\phi, \kappa) g_\lambda(\theta,\phi;\kappa) \cos\theta d\theta d\phi .
\]
For $\kappa=0$ and $\lambda=2$, the estimator \eqref{gkmodifieddef} reduces to the original G\&K estimator
defined in \eqref{gkcanestdef}.

\subsection{PARK normalized estimator}

The canonical PARK variance estimator is given by
\begin{equation}\label{parkoriginal}
\tilde{s}_\textrm{P} = {(H-L)^2 \over 4 \ln 2} .
\end{equation}
We generalize it by the corresponding normalized PARK estimator of order $\lambda$,
\begin{equation}\label{rsmodifieddef}
\tilde{e}_{\textrm{P},\lambda} = {R^\lambda \over
\mathcal{M}_{\textrm{P},\lambda}(\kappa)}~ \psi_\textrm{P}^{\lambda/2}(\Theta,\Phi) ,
\end{equation}
where
\[
\psi_\textrm{P}(\theta,\phi) = {\cos^2 \theta (1 - \sin 2 \phi) \over 4 \ln 2} , \quad \mathcal{M}_{\textrm{P},\lambda}(\kappa) = \iint\limits_{\mathcal{S}_\kappa} \psi_\textrm{P}^{\lambda/2}(\theta,\phi) g_\lambda(\theta,\phi;\kappa) \cos\theta d\theta d\phi .
\]

\begin{remark}
\textnormal{Below, we compare the efficiencies of the G\&K, of the PARK and of the most efficient estimators, and do not discuss the efficiency of the Roger-Satchell estimator. Indeed, it follows from our preliminary calculations for $\kappa\simeq 1$ that the  Rogers-Satchell bridge estimator is significantly less efficient than even the PARK estimator.}
\end{remark}

\subsection{Comparison of variance estimators}

Figure~3 shows the
dependence as a function of the bridge parameter $\kappa$ of the expected values of the G\&K \eqref{gkbridgest} and PARK \eqref{parkoriginal} variance estimators, in the case of zero drift ($\gamma=0$). One can observe that, for $\kappa\neq 0$, the G\&K and PARK variance estimators are biased, so it is convenient to compare
their normalized versions \eqref{gkmodifieddef} and \eqref{rsmodifieddef}.

\begin{figure}[h!]
\begin{center}
\includegraphics[width=0.7\textwidth]{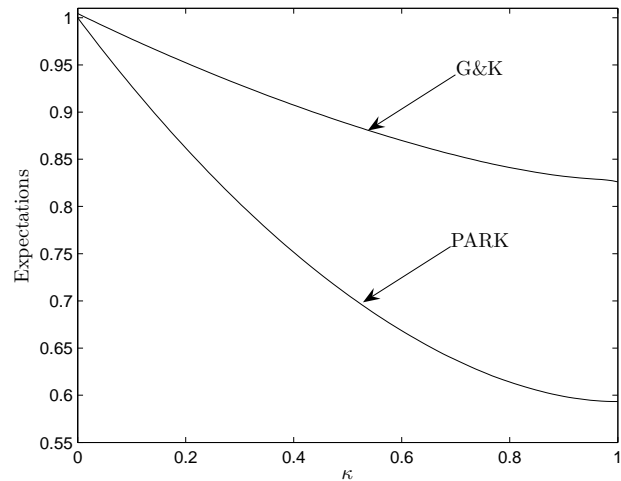}
\caption{Dependence of the expected values of the G\&K \eqref{gkbridgest}  and PARK \eqref{parkoriginal} canonical bridge estimators as a function of the parameter $\kappa$, in the zero drift ($\gamma=0$) case}
\end{center}
\end{figure}

\begin{figure}[h!]
\begin{center}
\includegraphics[width=0.7\textwidth]{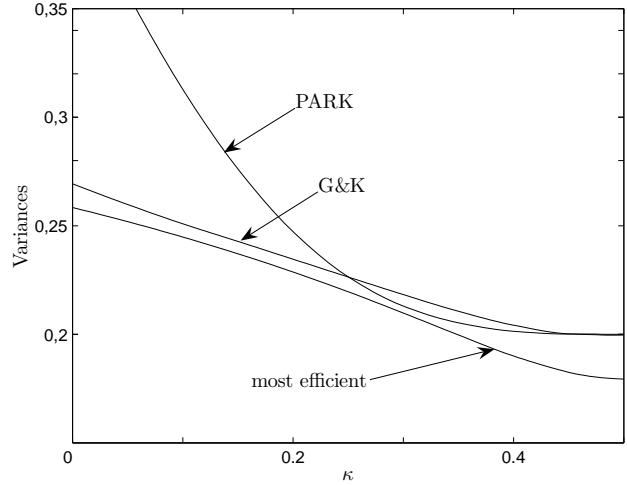}
\caption{Variances of the most efficient, normalized G\&K \eqref{gkmodifieddef} and PARK \eqref{rsmodifieddef} variance bridge estimators, as functions of the parameter $\kappa$, in the case of zero drift ($\gamma=0$)}
\end{center}
\end{figure}

Figure~4 plots the numerically calculated dependencies as a function of $\kappa$ of the variances of the most efficient canonical variance bridge estimator, with diagram \eqref{diagrmostefdef}, \eqref{mathedef} ($\lambda=2$), and of the variances of the G\&K and PARK canonical variance estimators \eqref{gkmodifieddef}, \eqref{rsmodifieddef}. For $\kappa=0$, i.e. in the case of ``standard'' OHLC estimators, the variances of the most efficient and of the G\&K estimators are rather close to each other, while the variance of the PARK estimator is much larger than the former ones:
\[
\textrm{Var}[\hat{e}_{\textrm{me},2}|\kappa=0]=0.2584 , \quad \textrm{Var}[\tilde{e}_{\textrm{GK},2}|\kappa=0]=0.2693 , \quad \textrm{Var}[\tilde{e}_{\textrm{P},2}|\kappa=0]=0.4073 .
\]
In contrast, in the case of an almost complete bridge $\kappa\in(0.9,1)$, the variance of the most efficient variance estimator is significantly smaller than the variances of the G\&K and PARK estimators:
\[
\textrm{Var}[\hat{e}_{\textrm{me},2}|\kappa=1]=0.1794 , \quad \textrm{Var}[\tilde{e}_{\textrm{GK},2}|\kappa=1] \simeq \textrm{Var}[\tilde{e}_{\textrm{P},2}|\kappa=1] \simeq 0.2 .
\]
Notice that the efficiencies of the G\&K and PARK estimators almost coincide for $\kappa\simeq 1$. This is due to the fact that, for $\kappa= 1$, the G\&K variance bridge estimator \eqref{gkbridgest} becomes close to the PARK estimator:
\[
\tilde{e}_{\textrm{P},2}(\kappa=1)\simeq \tilde{e}_{\textrm{GK},2}(\kappa=1) \sim k_1 (H-L)^2 + 2 k_2 HL, \quad k_1=0.511, \quad  k_2 = 0.0019 .
\]

\begin{figure}[h!]
\begin{center}
\includegraphics[width=1\textwidth]{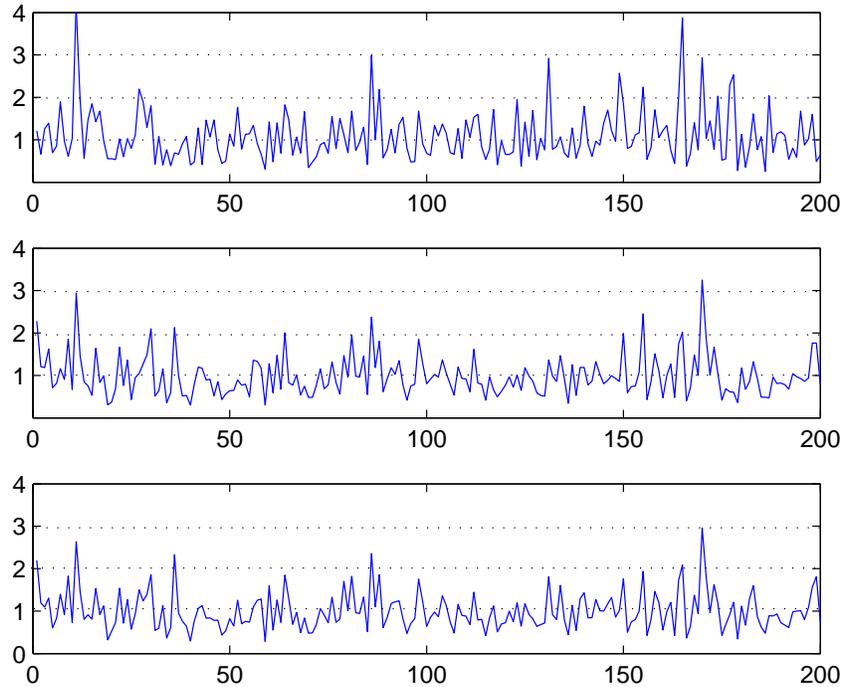}
\caption{Top to bottom: 200 samples of the G\&K estimator \eqref{gkcanestdef}, PARK estimator \eqref{rsmodifieddef} ($\lambda=2$) and most efficient canonical variance bridge estimator, for 200 realizations of a Wiener process  with zero drift ($\gamma=0$) and for $\kappa=0.99$}
\end{center}
\end{figure}

Simulating 200 realizations of a Wiener process and recording the associated OHLC,
figure~5 shows the  200 corresponding
G\&K estimator \eqref{gkcanestdef}, PARK estimator \eqref{rsmodifieddef} ($\lambda=2$) and
most efficient canonical variance bridge estimator, for $\kappa=0.99$ and $\gamma=0$.
It is visually apparent that the most efficient variance bridge estimator
exhibits smaller fluctuations and is more efficient than the PARK and G\&K bridge variance estimators.

\begin{figure}[h!]
\begin{center}
\includegraphics[width=0.7\textwidth]{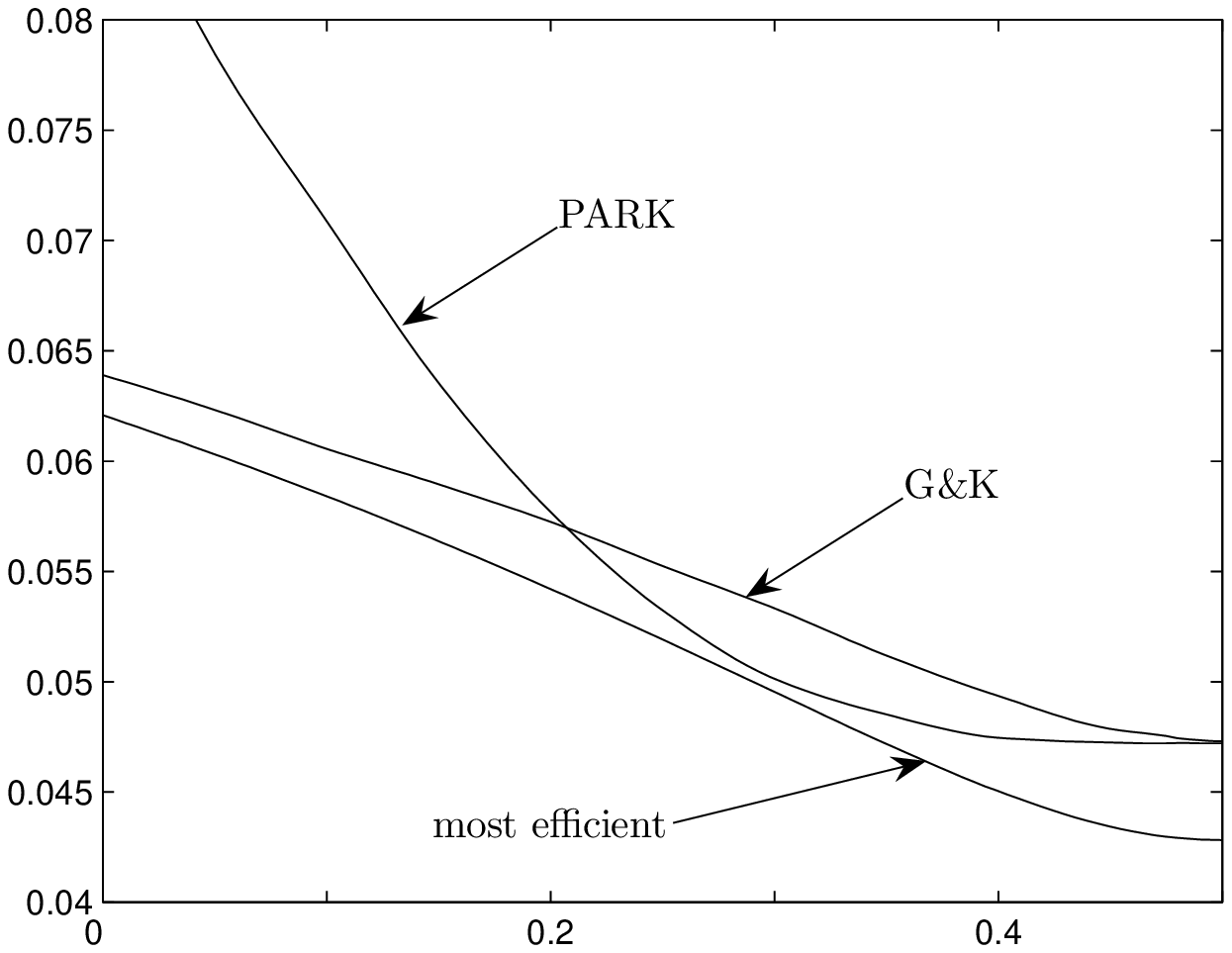}
\caption{Variances of the most efficient canonical volatility bridge estimator, G\&K \eqref{gkmodifieddef} and PARK \eqref{rsmodifieddef} volatility estimators ($\lambda=1$), as functions of $\kappa$ (for $\gamma=0$)}
\end{center}
\end{figure}
\begin{figure}[h!]
\begin{center}
\includegraphics[width=1\textwidth]{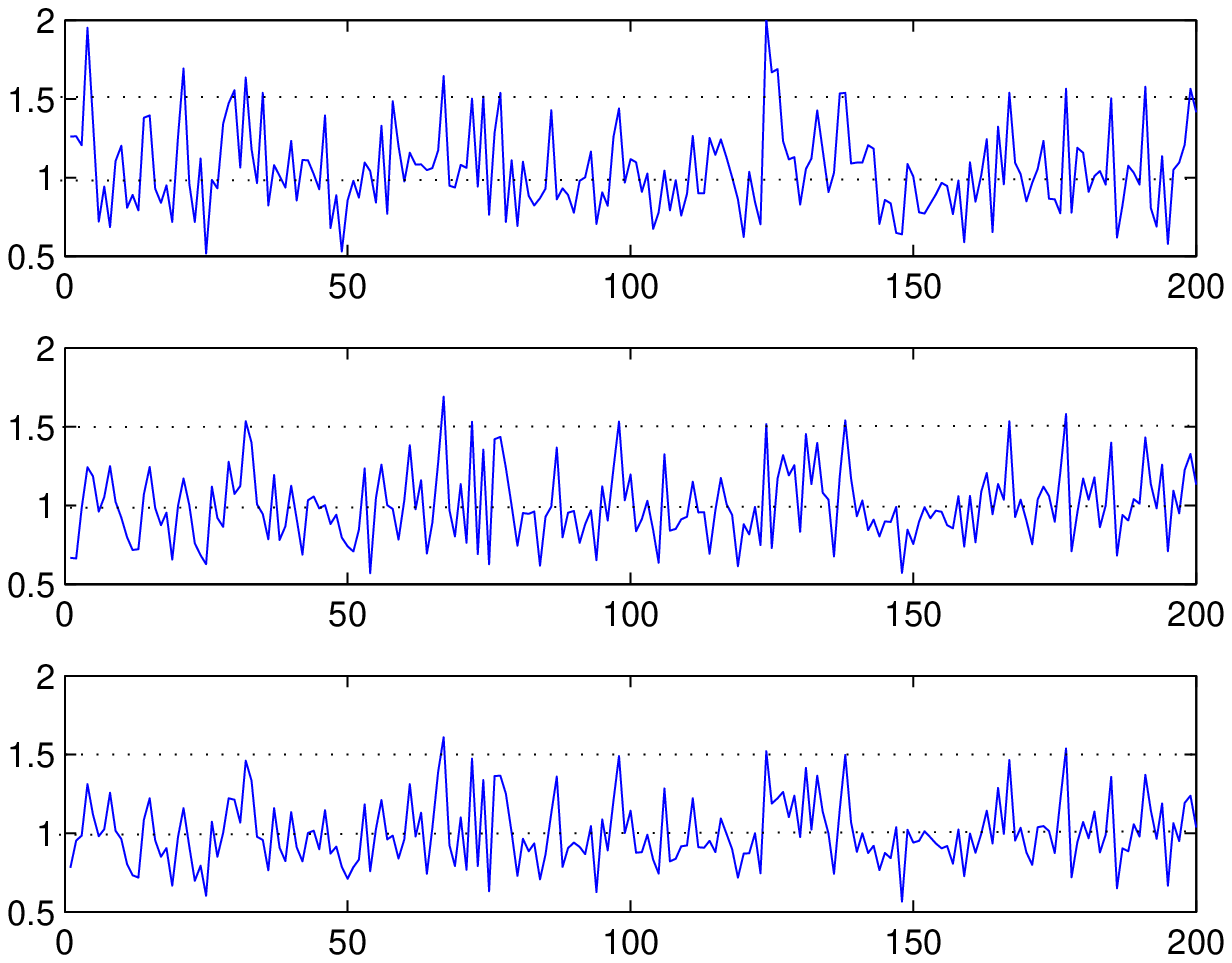}
\caption{Top to bottom: 200 samples of the G\&K volatility estimator, for $\kappa=0$, PARK volatility estimator and most efficient canonical bridge volatility estimator, for $\kappa=0.99$}
\end{center}
\end{figure}

\begin{remark}
\textnormal{The fact that the OHLC bridge estimators, with $\kappa\simeq 1$, are significantly more efficient than the ``standard'' OHLC estimators, corresponding to $\kappa=0$, can be intuitively explained as follows.
It is well-known that the high and low values of a Wiener process are most probably found in
the neighborhood of the edges of the observation interval. In contrast, by construction of the bridge, its high and low values are in general distant from the edges, as illustrated in figure~1. As a result, the high and low of a bridge incorporate significantly more information about the behavior of the original stochastic process than its own high and low values.}
\end{remark}

\begin{remark}
\textnormal{It is noteworthy that
the most efficient estimator at $\gamma =0$ remains
more efficient than the G\&K and PARK estimators
as long as $\gamma$ remains less than $0.8$ (for $\kappa= 0.95$)
and similar values for other $\kappa$'s. These condition are not
restrictive since relevant values of $\gamma$ are quite small.
Indeed, consider a typical stock with yearly volatility $\sigma =0.2$
and mean return $\mu=0.1$. Then, the value of $\gamma$ for
an estimator calculated at the daily scale $T \simeq 0.004$ year
is $\gamma = (\mu / \sigma) \sqrt{T} \simeq 0.032$. For estimators
at intra-day high-frequencies, for instance for $T= 5$ minutes $=0.00004$ year,
we have $\gamma \simeq 0.0032$. }
\end{remark}

\subsection{Comparison of volatility estimators}

Figure~6 shows the dependencies as a function of $\kappa$ of the variances of the most efficient canonical bridge volatility ($\lambda=1$) estimator, and the variances of the corresponding G\&K and PARK volatility estimators \eqref{gkmodifieddef}, \eqref{rsmodifieddef}. In the case of almost complete bridges $\kappa\in(0.9,1)$, the variance of the most efficient volatility estimator is significantly smaller than the variances of the analogous G\&K and PARK estimators:
\[
\textrm{Var}[\hat{e}_{\textrm{me},1}|\kappa=1]=0.0428 ,
\quad
\textrm{Var}[\tilde{e}_{\textrm{GK},1}|\kappa=1] = 0.0473 , \quad \textrm{Var}[\tilde{e}_{\textrm{P},1}|\kappa=1] = 0.0472 .
\]

Simulating 200 realizations of a Wiener process and recording the associated OHLC,
figure~7 shows the  200 corresponding G\&K volatility  ($\lambda=1$) estimator \eqref{gkmodifieddef}
for $\kappa=0$, the PARK estimator \eqref{rsmodifieddef}  ($\lambda=1$) and
the most efficient canonical bridge volatility estimators, for $\kappa=0.99$ and $\gamma=0$.
It is visually apparent that the most efficient volatility bridge estimator
exhibits smaller fluctuations and is more efficient than the PARK and G\&K bridge variance estimators.

\begin{figure}[h!]
\begin{center}
\includegraphics[width=0.99\textwidth]{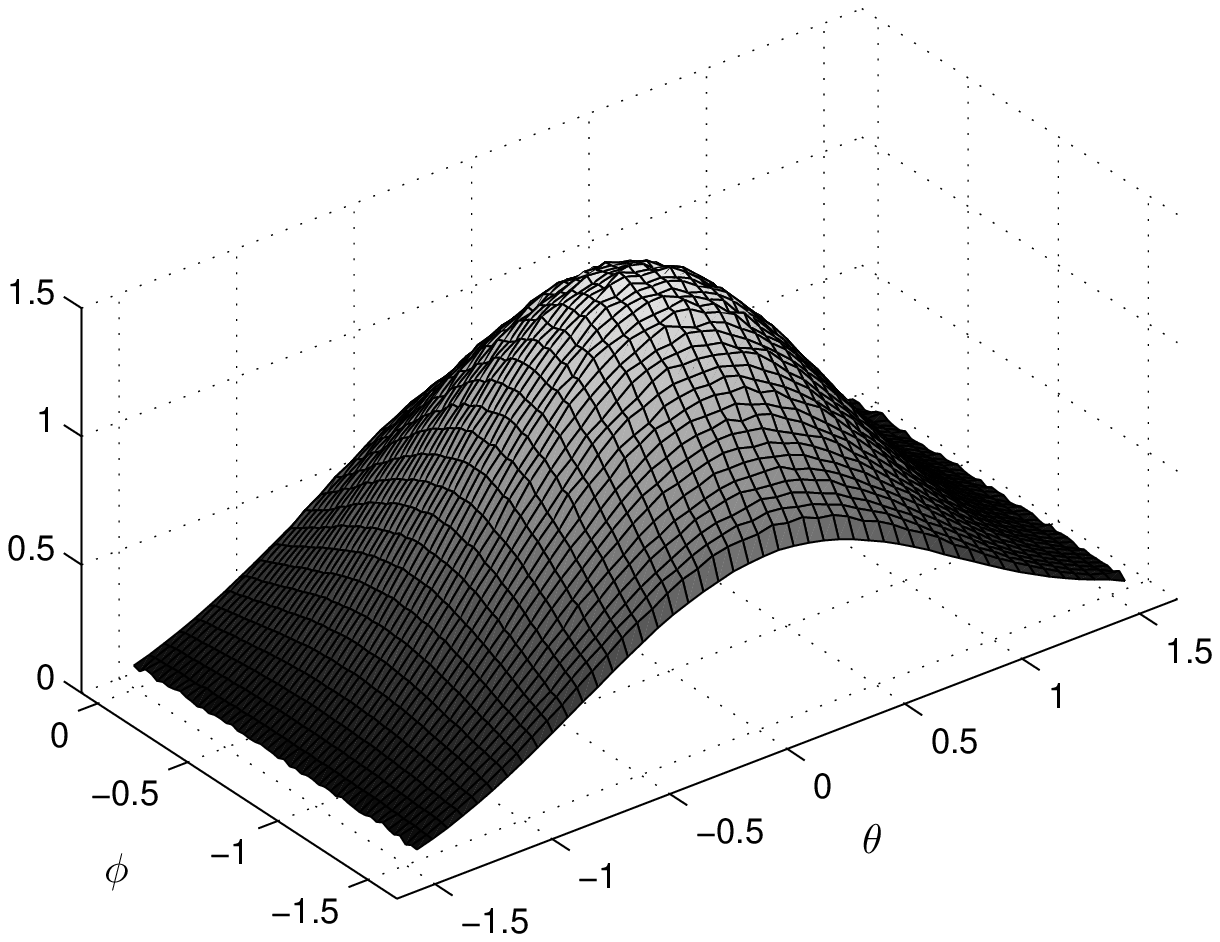}
\caption{Synthetic diagram of the most efficient variance canonical bridge estimator for $K=10$ and $\kappa=1$, $\gamma=0$}

\end{center}
\end{figure}
\begin{figure}[h!]
\begin{center}
\includegraphics[width=0.99\textwidth]{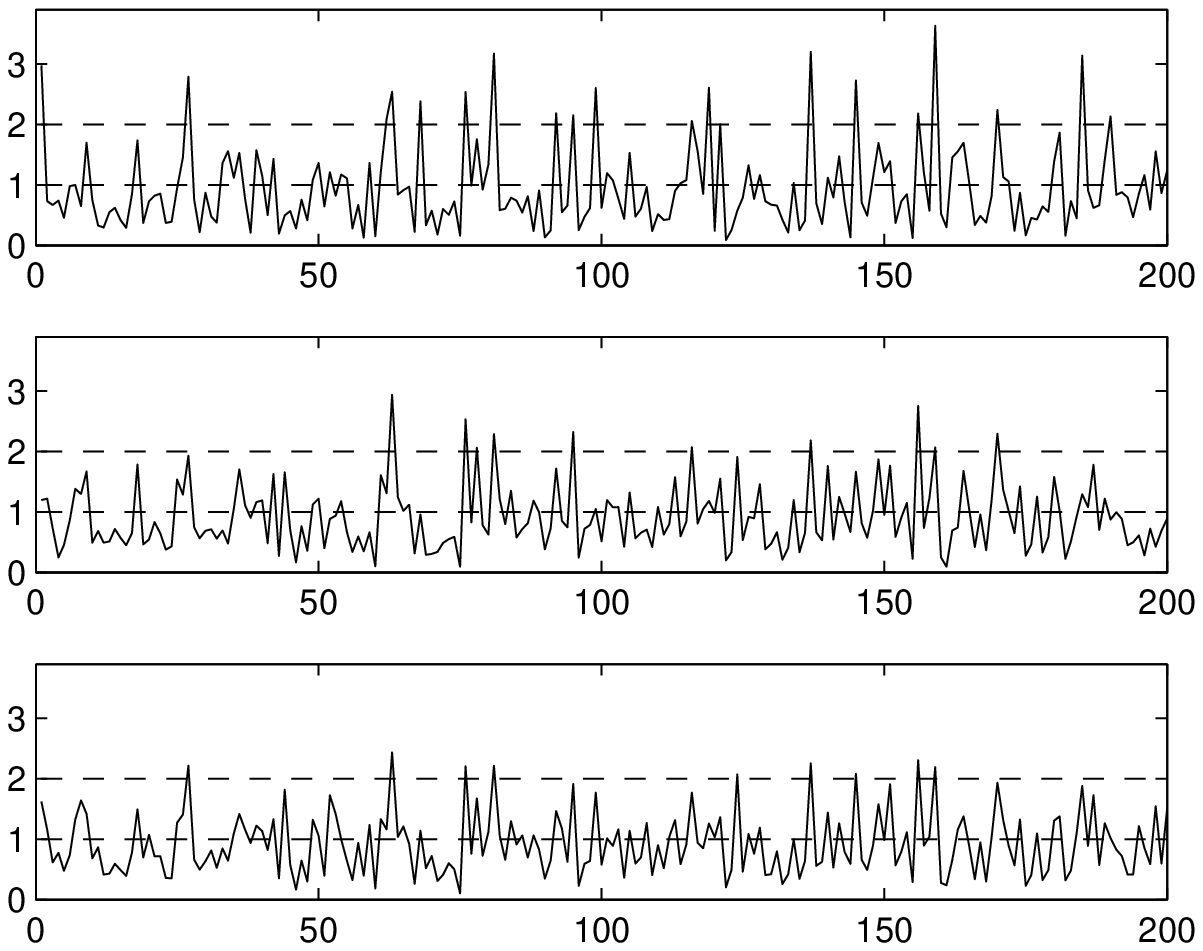}
\caption{Top to bottom: numerical samples of the G\&K estimator, for $\kappa=0$ and $\kappa=1$, and the numerical samples of the simulated most efficient canonical bridge variance estimator, for $\kappa=1$, obtained for discrete random walk $X(k)$ \eqref{xdiscrdef} with $K=10$}
\end{center}
\end{figure}

\section{Simulated most efficient estimators}

\setcounter{equation}{0}
\setcounter{theorem}{0}
\setcounter{remark}{0}
\setcounter{lemma}{0}
\setcounter{definition}{0}

The previous sections have derived the most efficient homogeneous bridge estimators, whose diagrams \eqref{diagrmostefdef}, \eqref{mathedef}, are defined in terms of the function $g_\lambda(\theta,\phi; \kappa, \gamma)$ \eqref{glambdef}, which depends in turn on the pdf $Q(h,\ell,c;\kappa,\gamma)$ \eqref{qhlcpdf}.  Notice that relation \eqref{glambdef} allows one to determine the function $g_\lambda(\theta,\phi; \kappa, \gamma)$ even
when the pdf is unknown. The function $g_\lambda(\theta,\phi; \kappa, \gamma)$ can indeed be
determined by simulating $M\gg 1$ times the stochastic process $X(t)$ which
describes the statistical properties of the log-price dynamics, and then estimate the function $g_\lambda(\theta,\phi; \kappa, \gamma)$ by its statistical average. This is particularly convenient when
theoretical formulas are not available, as occurs when considering stochastic processes
more complex than the Wiener process with drift.

To illustrate this possibility of simulating the diagrams associated with the most efficient estimators,
consider the discrete normalized random walk
\begin{equation}\label{xdiscrdef}
X(k) = {1 \over \sqrt{K}} \sum_{i=1}^k \epsilon_i , \qquad k=1,\dots, K, \qquad X(0)=0 ,
\end{equation}
where $\{\epsilon_i\}$ is a sequence of iid random variables with zero expectation and unit variance.
The random walk \eqref{xdiscrdef} mimics the discrete, tick-by-tick, nature of the log-price stochastic process.

\begin{table}[h!]
\caption{\label{tab01}Variances of G\&K and simulated most efficient variance estimators. The variances of the G\&K estimators and of the simulated most efficient variance estimators, both for $\kappa=0$ and $\kappa=1$, are obtained by averaging over $N=10^6$ simulations of the discrete random walk $X(k)$ \eqref{xdiscrdef}.}
\begin{center}
\begin{tabular*}{8.5cm}{@{}ccccc@{}}
\hline\hline
  & $K= 10$  & 100  & 1000  & $\infty$ \\
\hline
 $\textrm{Var}[\hat{e}_{\textrm{GK},2}](\kappa=0)$ & 0.5103 & 0.3272 & 0.2858 & 0.2693 \\
  $\textrm{Var}[\hat{e}_{\textrm{me},2}](\kappa=0)$   & 0.4759 & 0.3130 & 0.2755 & 0.2584 \\
$\textrm{Var}[\hat{e}_{\textrm{GK},2}](\kappa=1)$ & 0.4062 & 0.2434 & 0.2125 & 0.1996 \\
  $\textrm{Var}[\hat{e}_{\textrm{me},2}](\kappa=1)$ & 0.3373 & 0.2151 & 0.1896 & 0.1794 \\
  \hline\hline
\end{tabular*}
\end{center}
\footnotesize
\renewcommand{\baselineskip}{11pt}
\end{table}

In the limit $K\to\infty$, the random walk $X(k)$ \eqref{xdiscrdef} converges (even if $\{\epsilon_i\}$ are non-Gaussian as long as the tail of their pdf is not too heavy) to the Wiener process $W(t)$, so that the joint pdf of the random variables $\{H,L,C\}$ is known theoretically. In contrast, in the case of ``finite number of ticks'' ($K<\infty$), the joint pdf is unknown. Nevertheless, one can obtain an approximate expression for the diagram of the most efficient estimator by numerical simulation.

In order to construct the simulated diagram for $K=10; 10^2$ and $10^3$,  we divided the domain $S_\kappa$
defined in (\ref{thyjumldls}) in $50\times 50$ rectangle bins and, for each $K$, we generated $M=10^8$ simulations of the random walk $X(k)$ \eqref{xdiscrdef} with Gaussian summands $\{\epsilon_i\}$. We then calculated the function $g_\lambda(\theta,\phi;\kappa)$ (for $\gamma=0$) using the approximate statistical relation
\[
g_\lambda(\theta,\phi;\kappa) \cos\theta d\theta d\phi \simeq {1 \over M} \sum_{m=1}^M R_m^\lambda \textrm{I}_\delta (\Theta_m, \Phi_m) .
\]
The set $\{\Theta_m,\Phi_m,R_m\}$ are samples of the random variables \eqref{geogrback} obtained for the $m$-th simulation, $\textrm{I}_\delta$ is the indicator of the set
\[
\delta = (\theta,\theta+d\theta) \times (\phi,\phi+d\phi) ,
\]
and the summation is performed over $M$ simulations of the random walk \eqref{xdiscrdef}.
The histograms of the function $g_\lambda(\theta,\phi;\kappa)$ thus obtained is then substituted into the diagram function \eqref{diagrmostefdef}, \eqref{mathedef} to produce its 2D linear interpolation.

Figure~8 presents the 3D plot of the synthetic diagram of the most efficient variance estimator, obtained
by statistical averaging for $K=10$ and $\kappa=1$, $\gamma=0$.
Notwithstanding the visible fluctuations, table~1 shows that this level of numerical approximation
is sufficient to obtain significantly better efficient estimators than for instance, G\&K estimator.
Table~1 gives the variances of the canonical variance bridge estimators. The values shown in table~1
have been obtained by statistical averaging over $N=10^6$ simulations of the random walk \eqref{xdiscrdef}.

Simulating 200 realizations of a Wiener process \eqref{xdiscrdef} for $K=10$,
figure~9 shows the simulated most efficient canonical variance bridge estimator, for $\kappa=1$, and the samples of the G\&K estimators, for $\kappa=0$ and $\kappa=1$. It is clear that the simulated most efficient bridge variance estimator is significantly more efficient than the G\&K estimator.

\section{Conclusions}

In this paper, we have pursued the development of a comprehensive theory of homogeneous volatility estimators of
arbitrary stochastic processes. Our focus has been to derive OHLC (open-high-low-close)
log-prices bridge volatility estimators, which can span time intervals extending from seconds to years.
The main tool of our theory is the parsimonious encoding
of all the information contained in the OHLC in the form of general ``diagrams'' associated with the
joint distributions of the high-minus-open, low-minus-open and close-minus-open values of the original log-price process and its bridge. The diagrams can be tailored to yield the most efficient estimators associated to any statistical properties of the underlying log-price stochastic process.

Previous works have developed variance estimators
which are quadratic functions of the OHLC.
Our main contribution is to stress the remarkable fact that
quadratic estimators are only particular cases of general homogeneous
estimators. Our theory constructs the tools to find most efficient homogenous
estimators which, by construction, are always more efficient than the most
efficient quadratic estimators. Perhaps paradoxically, it turns out that the
search for the most efficient quadratic estimators is more tedious
than that of the more efficient homogeneous
estimators. Another advantage of homogeneous estimators is that they give
the possibility to develop efficient volatility in addition to variance estimators,
while quadratic estimators are specialized to variance estimators.

Our theory opens several interesting developments. First, the determination of the key functions $g_\lambda(\theta,\phi;\gamma)$, defining the above diagrams, provides the tools to develop efficient bridge volatility estimators for arbitrary non-Gaussian log-price processes, including the presence of
micro-structure as in tick-by-tick price series. Our methods should lead to the development of effective algorithms for low- and high-frequency OHLC volatility bridge estimators, that can be applied in practice to any kind of financial markets.

\section*{Acknowledgments}
One of us (FC) was inspired on the subject of this paper via an early collaboration with Prof. Curci.

\appendix
\setcounter{section}{0}
\setcounter{equation}{0}
\renewcommand{\theequation}{\thesection.{\arabic{equation}}}
\renewcommand{\thesection}{\Alph{section}}

\section{Appendix}

\subsection{Proof of Theorem \ref{mostefth} \label{thyh3q}}

For given values of the parameters $\kappa$ and $\gamma=\gamma_0$, the variance of the unbiased canonical estimator, with diagram \eqref{unbiasdiagr}, is equal to
\begin{equation}\label{varexpr}
\textrm{Var}[\hat{e}_\lambda;\kappa,\gamma_0] = {\iint\limits_{\mathcal{S}_\kappa}  G^2(\theta,\phi) ~ g_{2\lambda}(\theta,\phi;\kappa,\gamma_0) \cos \theta d\theta d\phi \over \left(~\iint\limits_{\mathcal{S}_\kappa} G(\theta,\phi) ~ g_\lambda(\theta,\phi;\kappa,\gamma_0) \cos \theta d\theta d\phi\right)^2} -1 .
\end{equation}
Using the Schwarz inequality
\[
\left(\iint\limits_{\mathcal{S}_\kappa} A(\theta,\phi) B(\theta,\phi)d\theta d\phi\right)^2
\leqslant \iint\limits_{\mathcal{S}_\kappa} A^2(\theta,\phi) d\theta d\phi \iint\limits_{\mathcal{S}_\kappa}
B^2(\theta,\phi) d\theta d\phi ,
\]
for arbitrary locally integrable real-valued functions $A(\theta,\phi)$ and $B(\theta,\phi)$,
we take
\begin{eqnarray}
A(\theta,\phi) &=& G(\theta,\phi)
\sqrt{g_{2\lambda}(\theta,\phi;\kappa,\gamma_0) \cos \theta} , \nonumber
\\
B(\theta,\phi) &=&
g_\lambda(\theta,\phi;\kappa,\gamma_0) \sqrt{{\cos \theta \over
g_{2\lambda}(\theta,\phi;\kappa,\gamma_0)}} , \nonumber
\end{eqnarray}
and obtain
\begin{eqnarray}
&& \left(~\iint\limits_{\mathcal{S}_\kappa}  G(\theta,\phi) g_\lambda(\theta,\phi;\kappa,\gamma) \cos\theta d\theta d\phi \right)^2 \leqslant \nonumber \\
&& \iint\limits_{\mathcal{S}_\kappa} G^2(\theta,\phi) g_{2\lambda}(\theta,\phi;\kappa,\gamma_0) \cos\theta
d\theta d\phi \iint\limits_{\mathcal{S}_\kappa} {g_\lambda^2(\theta,\phi;\kappa,\gamma_0) \over
g_{2\lambda}(\theta,\phi;\kappa,\gamma_0)} \cos\theta d\theta d\phi .
\nonumber
\end{eqnarray}
It follows from \eqref{varexpr} and from the above inequality that the variance of any canonical homogeneous volatility estimator of order $\lambda$ satisfies the inequality
\begin{equation}\label{varineq}
\textrm{Var}[\hat{e}_\lambda;\kappa,\gamma_0] \geqslant V_\lambda(\kappa,\gamma_0) , \qquad V_\lambda(\kappa,\gamma) = {1 \over \mathcal{E}_\lambda(\kappa,\gamma)} -1 ,
\end{equation}
where $\mathcal{E}_\lambda(\kappa,\gamma)$ is defined by expression \eqref{mathedef}. It follows from \eqref{varexpr}, \eqref{varineq} and \eqref{mathedef} that the variance of the canonical volatility estimator of order $\lambda$ reaches its minimal value for a given $\gamma=\gamma_0$ and $\kappa$, if $G(\theta,\phi)$ is given
by the left equality of  \eqref{mathedef}. \hfill $\blacksquare$

\subsection{Proof of Theorem \ref{jujuouyyn}  \label{thyh3qetgq}}

It is convenient to replace the initial condition \eqref{phiincond} by the more general one
\begin{equation}\label{phizerocond}
\varphi(\omega;\tau=0) = \varphi(\omega) , \qquad \omega\in(a,b) .
\end{equation}
At the end of proof, we obtain formula \eqref{solbidifprob} by taking $\varphi(\omega)=\delta(\omega)$.

The idea of the proof consists in redefining the function $\varphi(\omega)$ in \eqref{phizerocond} outside the interval $\omega\in(a,b)$ in such a way that the solution of equation \eqref{difeqphi}, supplemented by the initial condition
\begin{equation}\label{phizerocondinf}
\varphi(\omega;\tau=0) = \varphi(\omega), \qquad \omega\in(-\infty,\infty) ,
\end{equation}
satisfies the absorbing boundary conditions \eqref{absboundcond}. In other words, it should
be equal to zero on the lines
$\omega= a+\alpha \tau$, $\omega = b + \beta \tau$,  $\tau>0$.
Let us define the auxiliary function
\begin{equation}\label{phizerodef}
\varphi_0(\omega) = \varphi(\omega)\textrm{I}_{(a,b)}(\omega) , \qquad \omega \in (-\infty,\infty) ,
\end{equation}
where $\textrm{I}_{E}(x)$ is the indicator of the set $E$.

It follows from lemma~\ref{lem2} that the solution of the diffusion equation, supplemented by
the initial condition \eqref{phizerocondinf}, satisfies the boundary conditions \eqref{absboundcond} if $\varphi(\omega)$ obeys to symmetry relations
\begin{equation}\label{couplesymrel}
\varphi(\omega) = - \varphi(2 a - \omega)~ e^{2\alpha (a-\omega)} ,
\qquad \varphi(\omega) = - \varphi(2 b - \omega)~ e^{2\beta (b-\omega)} .
\end{equation}
Using the first of these two equalities and
definition \eqref{phizerodef} of the function $\varphi_0(\omega)$,
let us redefine $\varphi(\omega)$ onto the interval $\omega\in(2a-b,b)$ as follows:
\[
\varphi(\omega) = \varphi^0(\omega) , \qquad \omega\in (2a-b,b) ,
\qquad \varphi^0(\omega) =
\varphi_0(\omega) - \varphi_0(2a-\omega)~ e^{2\alpha(a-\omega)} .
\]
Then, the equalities \eqref{couplesymrel} provide the ``quasiperiodic'' relation
\[
\varphi(\omega) = \varphi(\omega+ 2(b-a)) ~ e^{2 (\beta-\alpha)(\omega +b-a) +2(\alpha b-\beta a)} ,
\]
which yields
\begin{equation}\label{phisum}
\varphi(\omega) = \sum_{m=-\infty}^\infty \varphi^m(\omega) ,
\end{equation}
where
\begin{equation}\label{phimsummanddef}
\varphi^m(\omega) = \varphi^0(\omega+2(b-a) m) ~
e^{2(\beta-\alpha)(\omega+ m \Delta) m +2(\alpha b -\beta a) m} .
\end{equation}
Substituting $\varphi(\omega)$ given by \eqref{phisum} with \eqref{phimsummanddef} into \eqref{soldifinconphi}, we obtain the sought solution of the initial-boundary problem \eqref{difeqphi}, \eqref{phiincond}, \eqref{absboundcond}. In particular, using $\varphi_0(\omega)=\delta(\omega)$, that is
\[
\varphi^0(\omega) \qquad \Rightarrow \qquad \varphi^0(\omega) =
 \delta(\omega) - e^{-2\alpha a}~\delta(\omega-2a) ,
\]
we obtain the solution \eqref{solbidifprob}. \hfill $\blacksquare$

\bibliographystyle{chicago}

\end{document}